\documentclass[preprint,12pt]{elsarticle}
\usepackage{latexsym, graphicx, epsfig, amsmath, amssymb, amsfonts}
\usepackage{amsmath}
\usepackage{amssymb}
\usepackage{graphicx}
\usepackage{subfig}
\usepackage{tabu}
\usepackage{adjustbox}
\usepackage{bbold}
\usepackage{bm}
\usepackage{textcomp}
\usepackage[thinc]{esdiff}
\usepackage[dvipsnames]{xcolor}
\usepackage{comment}
\usepackage{textgreek}
\usepackage{caption}
\usepackage{lineno}
\usepackage{multirow}
\usepackage{soul}
\usepackage{adjustbox}

\usepackage{siunitx}
\usepackage{xcolor}
\usepackage{booktabs}
\usepackage{caption}
\usepackage{float}
\usepackage{color,soul}

\journal{The Journal}
\begin{document}
\begin{frontmatter}

\title{Mechanical Characterization and Inverse Design of Stochastic Architected Metamaterials Using Neural Operators}

\author{Hanxun Jin\textsuperscript{a,d,}\fnref{firstfoot}}
\author{Enrui Zhang\textsuperscript{b,}\fnref{firstfoot}}
\author{Boyu Zhang\textsuperscript{a}}
\author{Sridhar Krishnaswamy\textsuperscript{a}}
\author{George Em Karniadakis\textsuperscript{b,c}}
\author{Horacio D. Espinosa\textsuperscript{a,}\corref{cor1}}

\fntext[firstfoot]{These two authors contribute equally}

\cortext[cor1]{Corresponding authors. E-mail:  espinosa@northwestern.edu (H.D.E.)}

\address[1]{Department of Mechanical Engineering, Northwestern University, Evanston, IL 60208}
\address[2]{Division of Applied Mathematics, Brown University, Providence, RI 02912}
\address[3]{School of Engineering, Brown University, Providence, RI 02912}
\address[4]{Division of Engineering and Applied Science, California Institute of Technology, Pasadena, CA 91125}

\begin{abstract}
Machine learning (ML) is emerging as a transformative tool for the design of architected materials, offering properties that far surpass those achievable through lab-based trial-and-error methods. However, a major challenge in current inverse design strategies is their reliance on extensive computational and/or experimental datasets, which becomes particularly problematic for designing micro-scale stochastic architected materials that exhibit nonlinear mechanical behaviors. Here, we introduce a new end-to-end scientific ML framework, leveraging deep neural operators (DeepONet), to directly learn the relationship between the complete microstructure and mechanical response of architected metamaterials from sparse but high-quality in situ experimental data. The approach facilitates the inverse design of structures tailored to specific nonlinear mechanical behaviors. Results obtained from spinodal microstructures, printed using two-photon lithography, reveal that the prediction error for mechanical responses is within a range of 5 - 10\%. Our work underscores that by employing neural operators with advanced micro-mechanics experimental techniques, the design of complex micro-architected materials with desired properties becomes feasible, even in scenarios constrained by data scarcity. Our work marks a significant advancement in the field of \textit{materials-by-design}, potentially heralding a new era in the discovery and development of next-generation metamaterials with unparalleled mechanical characteristics derived directly from experimental insights.
\end{abstract}


\end{frontmatter}


\newpage
\section{Introduction}

Designing material architectures at different scales, including truss-, plate-, or shell-based microstructures, unlocks unique properties distinct from their bulk counterparts, thereby significantly expanding the current material design space \cite{bauer2017nanolattices, zhang2020design, surjadi2019mechanical, schaedler2011ultralight, fleck2010micro,deshpande2001effective,jin2023mechanical}. For instance, periodic micro- and nano-lattices, fabricated from a wide variety of materials, exhibit extraordinary mechanical properties such as ultra-high specific strength \cite{meza2014strong, bauer2016approaching,zhang2019lightweight,jin2023ultrastrong}, energy absorption \cite{guell2019ultrahigh, frenzel2016tailored, bauer2022nanoarchitected}, and impact resilience \cite{portela2021supersonic}. Likewise, origami metamaterials exhibit foldability, auxeticity, tunable mechanical properties, and recoverability \cite{lin2020folding,lu2023origami,filipov2015origami}. These architected materials offer unprecedented flexibility in the design of structures with desired properties, which paves the way for designing the next-generation engineering materials \cite{bertoldi2017flexible, xia2022responsive}. Traditionally, such a design process has been intuition-based, often drawing inspiration from natural cellular materials, such as bones or nacres \cite{gibson2010cellular, espinosa2009merger, espinosa2011tablet, barthelat2007experimental, barthelat2011toughness}. More generally, the conceptualization of metamaterials with non-periodic microstructures, e.g., spinodal topology \cite{portela2020extreme,hsieh2019mechanical}, exhibit high mechanical resilience and tunable mechanical anisotropy. Interestingly, these microstructures were inspired by bi-continuous phases, e.g., those observed in bones and the spinodal decomposition emergent during copolymer synthesis \cite{khandpur1995polyisoprene, jin2022dynamic, jin2022big, jin2022understanding}. However, such an intuition-based design approach suffers from an inherently limited design space.

In this context, the emergence of machine learning (ML) tools has been promising because they offer new approaches for extracting insights from both experimental and computational data and pave the way for a novel approach in inverse design \cite{jin2023recent, zheng2023deep, deng2022inverse, kumar2020inverse, alderete2022machine, mao2020designing, ma2022deep,ha2023rapid, chen2020generative, bastek2022inverting}. Facilitated by ML, the design process is gradually steering from a heuristic process towards a more systematic and data-driven approach. However, this pivotal transition is contingent upon the availability of sufficient training datasets, typically sourced from computations such as finite element analysis (FEA) and large-scale molecular dynamics (MD). Moreover, the approach poses significant challenges when applied to complex material architectures due to the computational complexity of non-periodic structures and their multiscale material/structural responses, including nonlinearity, instabilities, and damage. Indeed, the inherent complexity of stochastic architected materials and their associated responses results in a scarcity of robust and reliable computational training datasets for ML applications, thereby creating a bottleneck for progress. Furthermore, defects and imperfections created during the fabrication process and the dependency of mechanical properties on process parameters mean that ML predictions can notably deviate from actual experimental outcomes \cite{jin2020machine, wang2020machine}. Therefore, the inverse design of these materials has been limited primarily to simple design criteria like homogenized elastic moduli \cite{kumar2020inverse, zheng2021data, senhora2022optimally}. Recently, an approach for inverse design of metamaterials beyond linear elasticity was advanced, which involves training an ML model directly from experimental datasets \cite{ha2023rapid}. While compelling, this approach has the potential drawback that it requires copious training datasets, with the associated fabrication and characterization burden. While some examples of autonomous experimentation have been reported for designing macro-scale metamaterials \cite{stach2021autonomous,gongora2020bayesian}, the field is still in its infancy, especially for conducting sophisticated in situ experiments needed for nano- and micro-architected material design. Therefore, a diversity of novel approaches for inverse design of multifunctional materials with complex architectures and desirable properties beyond elasticity, e.g., nonlinear responses associated with material nonlinearities, instabilities, and damage, are highly needed.

\begin{figure}[!ht]
    \centering
    \includegraphics[width=1.0\textwidth]{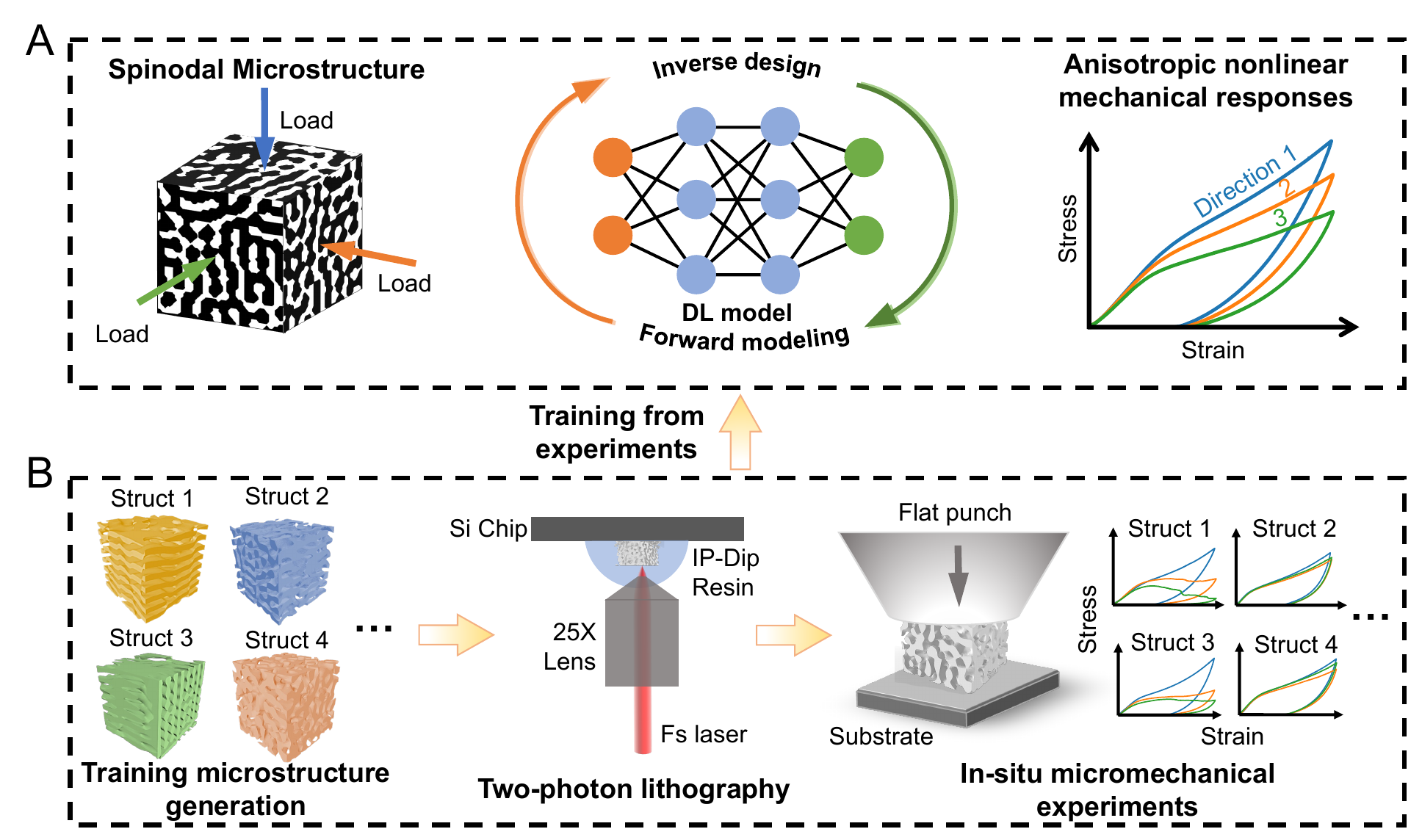}
    \caption{\textbf{Forward Modeling and Inverse Design of Anisotropic Spinodal Microstructures using Deep Learning based on Limited and Sparse in situ Micro-compression Experimental Data.} (A) Schematic of forward modeling and inverse design; (B) Training data collection from in situ micro-compression experiments on samples fabricated by two-photon lithography.}
    \label{fig:schematics}
\end{figure}

Herein, we address these challenges by introducing an innovative, end-to-end scientific machine learning (SciML) framework \cite{karniadakis2021physics} based on the deep neural operator (DeepONet) \cite{lu2021learning}, which can effectively design metamaterials based on sparse and limited experimental data (\textbf{Fig.~\ref{fig:schematics}A}). Despite our framework being scale-agnostic, enabling the design of architected materials across multiple scales, we focus on the inverse design of micro-architected spinodoid metamaterials as a proof of concept, targeting desired mechanical responses beyond linear elasticity. We collected a sparse yet high-quality experimental dataset from in situ micro-compression experiments conducted inside a scanning electron microscope (SEM) on structures fabricated through two-photon lithography (TPL) (\textbf{Fig.~\ref{fig:schematics}B}). Our results demonstrate that the proposed modified DeepONet is a promising surrogate model to predict the mechanical responses of micro-architected materials with an accuracy of 5 - 10\%. We also introduce a successful implementation of inverse design for desired stress-strain behaviors while avoiding unreliable and expansive computations. We anticipate that our methodology will lay the foundation for developing next-generation materials with tailored properties and functionalities when only limited data is available. 

\section{Results}

\subsection{Mechanical Data Collection from In situ Micromechanical Experiments}

\begin{figure}[!ht]
    \centering
    \includegraphics[width=1.0\textwidth]{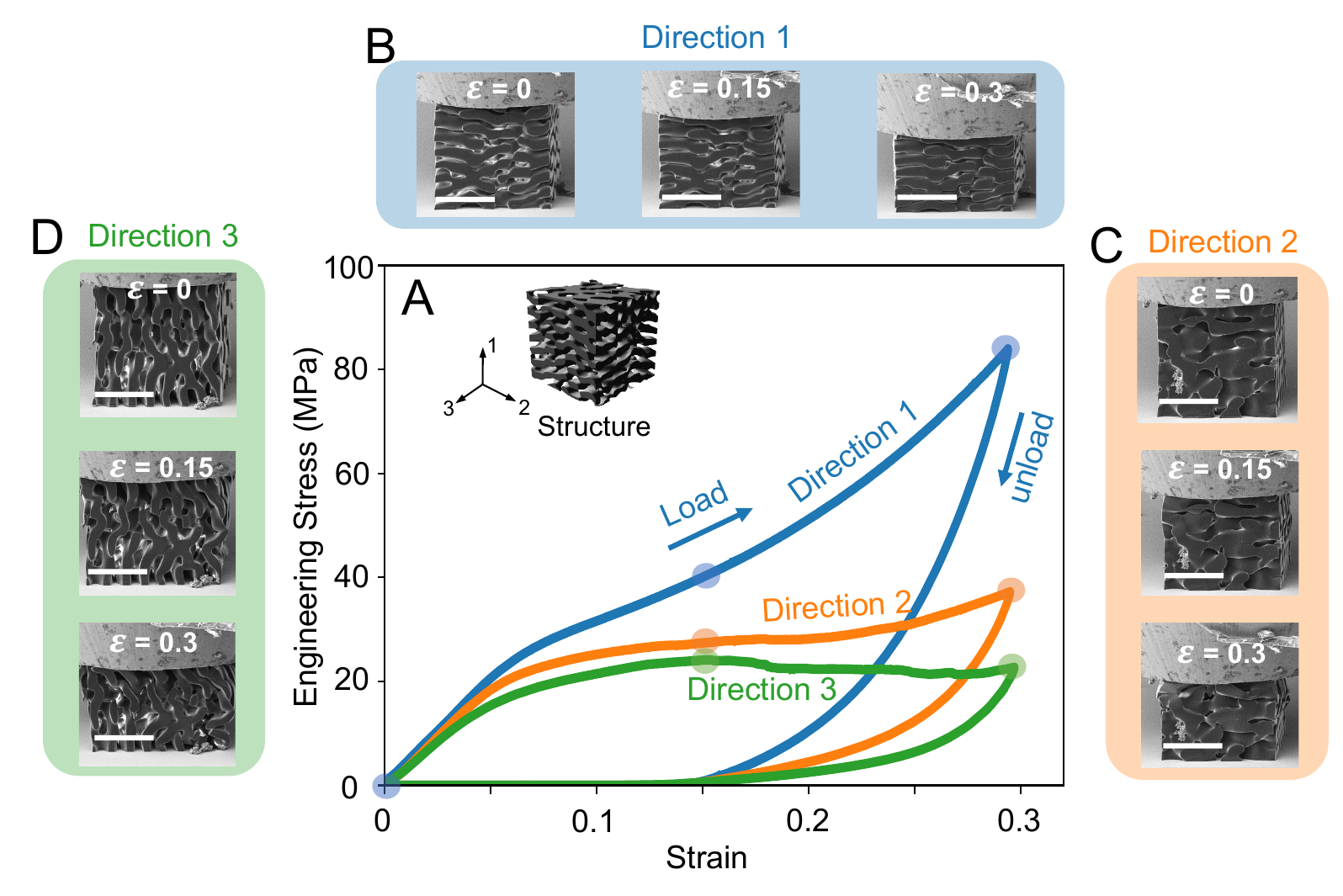}
    \caption{\textbf{In situ Micro-compression Experiment for a Spinodal Microstructure}. (A) Nonlinear and anisotropic stress-strain curves were observed in three principal directions. Insets show the microstructure design and Cartesian coordinates; (B) SEM images of the sample deformed in direction 1 at three different strain levels; (C) SEM images of the sample deformed in direction 2 at three different strain levels; (D) SEM images of the sample deformed in direction 3 at three different strain levels. The scale bar is 50 $\mu$m.}
    \label{fig:experiment}
\end{figure}

The mechanical behavior of spinodal metamaterials, such as stiffness and energy absorption, can be precisely adjusted through their microstructural design \cite{portela2020extreme}. These structures often exhibit marked anisotropy in their mechanical responses, meaning different mechanical responses when loaded in 3 principal directions. \textbf{Fig.~\ref{fig:experiment}A} illustrates this phenomenon with a lamellar-type microstructure fabricated using TPL, showcasing pronounced mechanical anisotropy. When the microstructure was loaded in direction 1, strain-hardening was evident when the strain exceeded 0.1. We attribute such behavior to the densification of the lamellar microstructure (\textbf{Fig.~\ref{fig:experiment}B} and \textbf{Movie S1}). In contrast, loading in directions 2 and 3 displayed negligible strain-hardening owing to structural buckling and crack formation (\textbf{Fig.~\ref{fig:experiment}C} \& \textbf{D}, and \textbf{Movie S2} \& \textbf{S3}). 

It is also worth noting that the mechanical properties can be largely influenced by the TPL fabrication parameters, such as laser power, slice/hatch distances, and the choice of photoresists. Consequently, our objective is to predict and design microstructures that exhibit specific mechanical behaviors while maintaining a consistent fabrication process and using the same base material. To this end, we performed in situ micro-compression experiments on 18 representative micro-scale spinodal structures fabricated via TPL, as listed in \textbf{Fig. S1}. These tests provided a sparse yet high-quality dataset of stress-strain curves for training the DeepONet. The experimental procedures are presented in \textbf{Materials and Methods}. Details regarding the generation of spinodal microstructures are presented in \textbf{Supplementary Materials}.

\subsection{DeepONet for Relating Microstructures with Nonlinear Mechanical Properties}

\begin{figure}[!ht]
    \centering
    \includegraphics[width=0.85\textwidth]{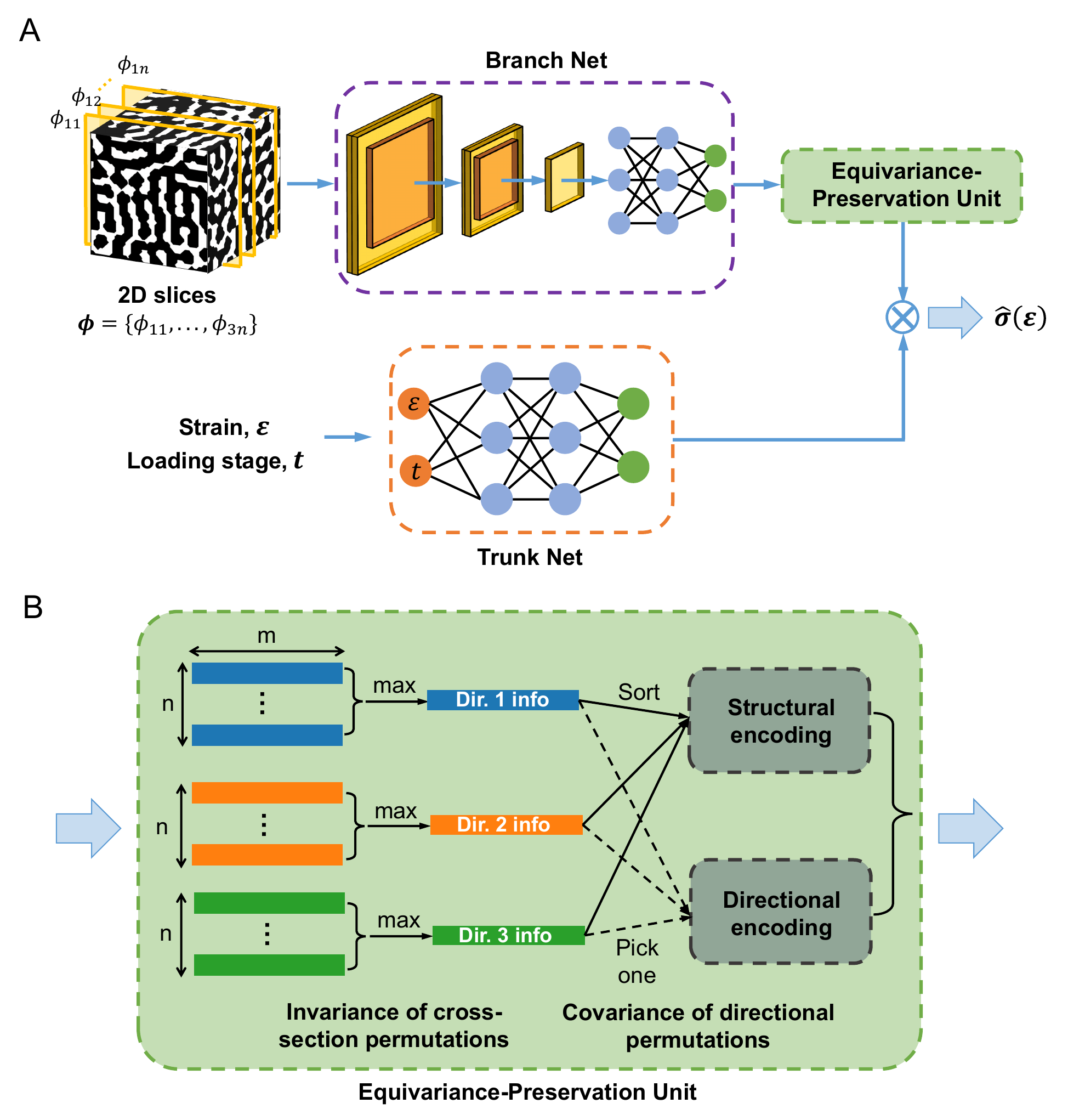}
    \caption{\textbf{DeepONet Structure for Relating Microstructures with Stress-strain Curves of Spinodal Materials.} (A) Schematic of the DeepONet architecture. The microstructure is preprocessed into 2D cross-sections, denoted as $\{\phi^{(pq)}\}_{p,q=1,1}^{3,n}$. These cross-sections are sequentially analyzed by the branch net and the equivariance-preserving unit. The loading strain $\varepsilon$ and the pseudo-time $t$ (indicating the loading/unloading step) are fed into the trunk net. The final output is the predicted stress $\hat{\sigma}$ corresponding to the input microstructure for the given strain $\varepsilon$ and loading step $t$. (B) Detailed structure of the equivariance-preserving unit; $m$ represents the dimension of the output vector from the branch net for each cross-section. This unit is designed to maintain invariance to permutations of cross-sections and covariance to directional permutations. This is achieved through max pooling for cross-section permutations (on the left) and structural and directional encoding for directional permutations (on the right).}
    \label{fig:NN}
\end{figure}

In our study, we employed DeepONet \cite{lu2021learning} to develop a SciML model tailored for analyzing the mechanical properties of materials with spinodal microstructures. DeepONet is an innovative architecture of neural networks specifically designed to learn (nonlinear) operators, i.e., mappings from functions to functions \cite{goswami2022physics}. Since its invention \cite{lu2021learning}, DeepONet has been successfully applied as surrogate models for various engineering problems, including multiscale mechanics \cite{yin2022interfacing,goswami2022physics}, phase transitions \cite{oommen2022learning,lin2021operator}, and biomechanics \cite{zhang2022g2varphinet, yin2023generative, yin2022simulating}. Here, the application of DeepONet focuses on analyzing the complex relationship between the 3D microstructures and their corresponding nonlinear stress-strain behaviors. We briefly summarize the DeepONet architecture (see \textbf{Fig.~\ref{fig:NN}}) in this section and provide complete technical details in \textbf{Materials and Methods} and \textbf{Supplementary Materials}.

The DeepONet architecture is shown in \textbf{Fig.~\ref{fig:NN}A}. Briefly, it maps the microstructural information into the corresponding stress-strain curves for a complete loading-unloading cycle. The DeepONet consists of a branch net and a trunk net, as proposed in \cite{lu2021learning}. Herein, we modify the basic architecture to endow it with an additional equivariance-preserving unit that accommodates the inherent symmetries specific to our study. The branch net takes the individual 2D cross-sections of the 3D microstructures as inputs, represented as $\{\phi^{(pq)}\}_{p,q=1,1}^{3,n}$, where $p$ indicates the normal direction, $q$ the index of the cross-section, and $n$ the total number of cross-sections. The trunk net inputs include the loading strain $\varepsilon$ and the pseudo-time $t$, which indicates whether the material is in the loading or unloading step. An equivariance-preserving unit is appended after the branch net. The final output of DeepONet is the stress prediction $\hat{\sigma}$ corresponding to the input microstructure $\{\phi^{(pq)}\}_{p,q=1,1}^{3,n}$, evaluated at the particular strain $\varepsilon$ and loading stage $t$. This stress prediction is derived by performing a ``dot product" of the output vectors from the equivariance-preserving unit and the trunk net, which have a dimensionality of $p$.

The equivariance-preserving unit, illustrated in \textbf{Fig.~\ref{fig:NN}B}, is designed to incorporate the inherent symmetries into the DeepONet framework. The input of this unit consists of $3n$ vectors, each of dimension $m$, originating from the branch net. We consider two types of symmetries, namely, the invariance of cross-section permutations and the covariance of directional permutations. The invariance of cross-section permutations means that permuting the sequence of the $n$ cross-sections within the same normal direction should not affect the resulting output stress-strain curve. While this does not hold precisely, it is a practical simplification adopted in our study to reduce the complexity and dimensionality of the input space. Although allowing the shuffling of these $x-y$ cross-sections, for example, disrupts the original sequential order of the cross-sections along the $z$ direction, we rely on $x-z$ and $y-z$ cross-sections to retrieve the sequential information along $z$ direction. The covariance of directional permutations implies that permuting the three spatial directions of the microstructure should result in a corresponding permutation in the output stress-strain curves. The implementation of these two types of symmetries is achieved through max pooling and structural and directional encoding. Max pooling, depicted on the left side of \textbf{Fig.~\ref{fig:NN}B}, addresses the invariance of cross-section permutations, while the structural and directional encoding, shown on the right side, deals with the covariance of directional permutations. Further details on this equivariance-preserving unit can be found in the \textbf{Supplementary Materials}.

To train and evaluate the modified DeepONet, we used the experimental stress-strain data as well as the augmented microstructure dataset. The augmented microstructure dataset comprises the microstructures actually printed experimentally as well as those variants with similar patterns that are generated with the same set of phase angle parameters (see \textbf{Supplementary Materials}). We assumed that microstructures created with identical phase angle parameters would exhibit the same mechanical behaviors. We trained the DeepONet using the experimental data over 20,000 epochs. We observed that both the training and testing losses decayed rapidly, reaching a plateau at approximately 10,000 epochs, as depicted in \textbf{Fig. S2}, which indicated that DeepONet can well-capture the relationship between the spinodal microstructure and the nonlinear, history-dependent stress-strain curves.

\subsection{Predicting Mechanical Responses of Unseen Micro-architectures}

\begin{figure}[!ht]
    \centering
    \includegraphics[width=1.0\textwidth]{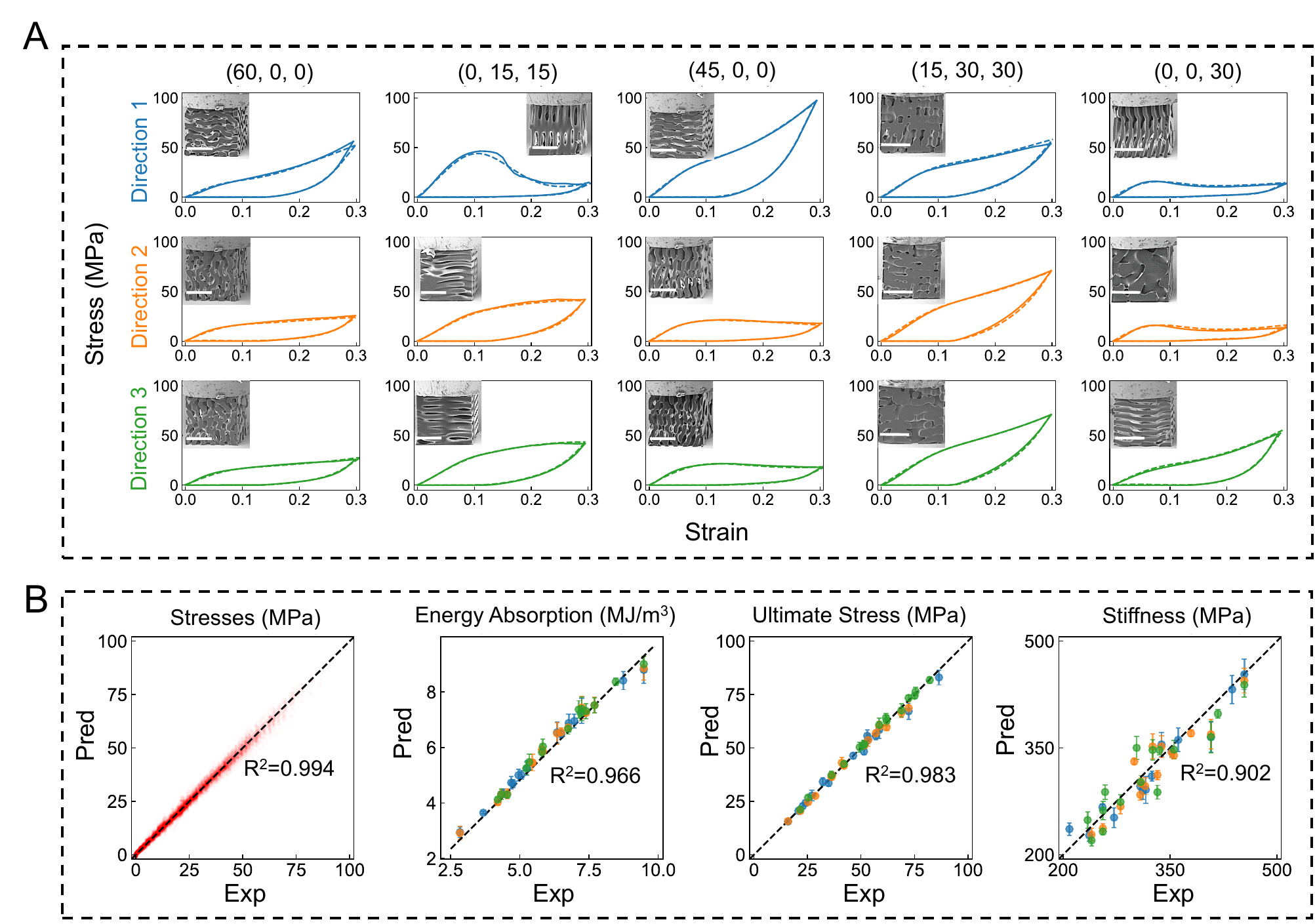}
    \caption{\textbf{Performance of the Trained DeepONet in Predicting Mechanical Properties of Unseen Microstructures.} (A) The experimental data and the predicted stress-strain curves of five typical spinodal microstructures. The solid lines represent the experimental data, and the dashed lines represent the DeepONet predictions. Insets show the corresponding spinodal structure. The corresponding phase angles are indicated above their respective panels. The scale bar is 50 $\mu$m. (B) Comparison of the predictions and the experimental values in terms of stress values, energy absorption, ultimate stress, and stiffness.}
    \label{fig:results_forward}
\end{figure}

We first assess the capability of the trained DeepONet to predict the mechanical properties of microstructures that the algorithm had not previously encountered. The predictions for the stress-strain curves along the three principal directions are presented in \textbf{Fig.~\ref{fig:results_forward}A}. In each case, we compare the experimental data (solid lines) with the DeepONet predictions (dashed lines). The DeepONet demonstrates remarkable accuracy in predicting both the loading and unloading stress-strain curves in all three directions, achieving a mean square error (MSE) of less than 2\%. This level of precision is particularly noteworthy given the limited experimental data available for training. Notably, the DeepONet not only accurately predicted strain hardening but also adeptly captured strain-softening behaviors resulting from structural buckling and failure, demonstrating its capability of handling highly nonlinear material behavior. Notable examples include the prediction for direction 1 of the (0, 15, 15) sample and direction 1 of the (0,0,30) sample.

To provide a quantitative evaluation of DeepONet's prediction performance, we present a detailed comparison of predicted versus experimental values for all stresses, energy absorption, ultimate stress, and stiffness across all test cases in \textbf{Fig.~\ref{fig:results_forward}B}. The predictions for all stresses, energy absorption, and ultimate stress show high accuracy, with a coefficient of determination ($R^2$) over 0.96. The $R^2$ value for stiffness is 0.90, which is lower than other quantities. This feature can be attributed to two main factors: (1) stiffness is calculated as the derivative of stress with respect to strain, a process that tends to amplify errors, and (2) the DeepONet architecture and its loss function are primarily tailored to model the overall nonlinear stress-strain relationship rather than focusing specifically on stiffness. For applications where accurate stiffness prediction is critical, modifications can be implemented, such as increasing weights on stress data in the linear region within the loss function. Despite these nuances, DeepONet demonstrates remarkable efficacy in predicting the mechanical properties of unseen microstructures, highlighting its potential as a robust tool for analyzing mechanical behaviors in various architectures, especially in situations where experimental data availability is limited.

\subsection{Inverse Design and Experimental Validation}

\begin{figure}[!ht]
    \centering
    \includegraphics[width=0.85\textwidth]{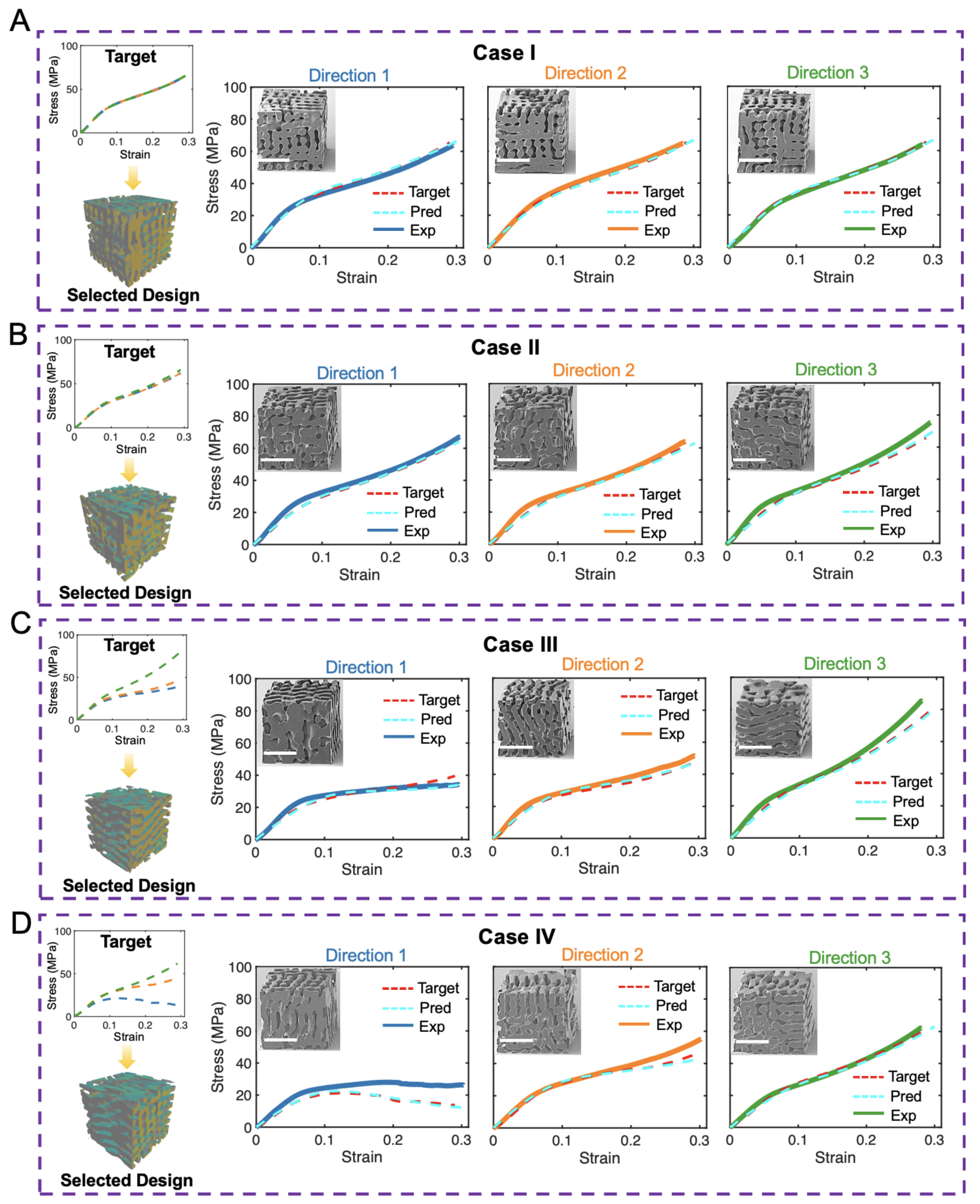}
    \caption{\textbf{Results of the Trained DeepONet in the Inverse Design of Spinodal Microstructures with Desired Mechanical Responses.} (A) Case I, (B) Case II, (C) Case III, and (D) Case IV. We conducted experimental validations for four distinct cases, each demonstrating different mechanical behaviors. Displayed are both the target stress-strain curves and those predicted by DeepONet, alongside the experimental data, for comparison in three  directions. Insets are SEM images for designed structures. The scale bar is 50 $\mu$m.}
    \label{fig:results_inverse}
\end{figure}

We have successfully demonstrated the efficacy and accuracy of the DeepONet model, primarily as a surrogate for forward modeling. However, the capabilities of DeepONet extend well beyond the forward modeling. In this section, we delve into the potential of our method for the inverse design of microstructures tailored to specific mechanical properties. While a comprehensive inverse design process involves forward simulation/modeling, (inverse) optimization algorithms, generation of candidate designs, defining performance criteria and objectives, and iterative feedback, we herein utilized a streamlined approach consisting of an indirect inverse design methodology. It is crucial to recognize that other inverse design methodologies could be effectively integrated into the proposed framework. For instance, employing genetic algorithms (GA) \cite{holland1992genetic} or leveraging generative ML methodologies such as (conditional) generative adversarial networks (GANs) \cite{goodfellow2014generative, mirza2014conditional}, can be employed to enable the inverse design of new micro-architectures, potentially expanding beyond those currently represented in the experimental database. The framework of the inverse design workflow is illustrated in \textbf{Fig. S3}. Prior to the design process, we obtained a dataset of 2,000 microstructures generated through the random sampling of phase angles. We then employed the trained DeepONet as a forward solver to predict the mechanical properties of these microstructures and then selected the design that most aligned with the desired stress-strain curves, as determined by the MSE metric. This process exclusively involves the forward propagation through DeepONet, enabling its completion within seconds.

We provided four representative stress-strain responses, as targets, which were not included in the DeepONet training data. These target curves were generated by mixing and/or averaging the curves from two experimental samples. After identifying the best design, we compared three sets of stress-strain curves: (1) the target stress-strain curves, (2) the DeepONet-predicted stress-strain curves for the designed microstructures, and (3) the experimental stress-strain curves of the designed microstructures. 

The outcomes for these four cases are displayed in \textbf{Fig.~\ref{fig:results_inverse}}, which include the target curves, the chosen designs, the fabricated microstructures, and a comparison of the three sets of curves. In \textbf{Case I}, our objective was to design a microstructure with isotropic mechanical responses (\textbf{Fig.~\ref{fig:results_inverse}A}). We identified the best design as the spinodal microstructure having an MSE of 1.07\% (averaged in 3 directions) between predicted and target stresses-strain behaviors. Experimental verification revealed an error of 3.14\% compared to the target stress-strain behavior. Considering the existence of stochastic fabrication defects and instrumentation errors during testing, the result represents a remarkable accuracy for designing isotropic spinodal microstructures. \textbf{Movies S4 - S6} display the in situ SEM experimental verification for \textbf{Case I} under three principal directions, respectively. For \textbf{Case II}, we introduced a slight variation in the mechanical response along direction 3, while keeping the responses in directions 1 and 2 identical (\textbf{Fig.~\ref{fig:results_inverse}B}). This subtle deviation led the inverse design algorithm to select a microstructure visually distinct from \textbf{Case I}. The MSE between the target and predicted stress-strain behaviors was 1.25\%, and the MSE between the experimental and predicted stresses was 6.0\%, again demonstrating high accuracy. In \textbf{Case III}, our goal was to design a microstructure with distinct anisotropy (\textbf{Fig.~\ref{fig:results_inverse}C}). The inverse design algorithm selected a lamellar-type microstructure, exhibiting moderate strain hardening in directions 1 and 2, due to lamellar buckling and failure, but a much prominent strain hardening in direction 3. Impressively, the predictions captured these complex behaviors with an MSE of 3.87\%, and the experimental stress-strain behavior presented an error of 7.61\%. These small errors highlight the algorithm's capability to inverse design microstructures with desired anisotropic properties. \textbf{Case IV} focused on designing a microstructure exhibiting distinct strain softening in one direction due to structural bulking, and hardening in the other two directions (\textbf{Fig.~\ref{fig:results_inverse}D} and \textbf{Movies S7 - S9}). A comparison between the stress-strain predictions for the selected design and experimentally measured behavior, showed MSEs of 7.99\% and 2.73\% in directions 2 and 3, respectively. The MSE in direction 1 was substantially higher, at 25.2\%, though the experimental results confirmed the predicted strain-softening behavior. We attribute the larger error in direction 1 to the role of imperfections in compressive buckling behaviors, rather than the limited training data obtained from strain-softening cases. This assertion is based on the fact that the DeepONet algorithm is capable of accurately predicting strain-softening behaviors in previously unseen microstructures (\textbf{Fig. 4}).

\section{Discussion}

Collecting high-fidelity experimental data from in situ micro- and nano-scale mechanical tests is a resource-intensive task, both in terms of time and cost. However, this data collection remains the most reliable method for obtaining precise mechanical information at small scales. This necessity underscores the importance of complementary SciML methods that can effectively build the surrogate models, particularly given the challenges posed by data scarcity as well as inherent nonlinearities and softening in mechanical responses.

In this work, we introduced a SciML approach specifically tailored to learn from limited experimental data on the anisotropic, nonlinear mechanical properties of micro-architected spinodal materials. To achieve the desired functionality, we carefully designed the SciML framework and processed the experimental data with several key strategies. First, we chose DeepONet as the base model, known for its efficacy as a nonlinear operator approximator and strong generalization capabilities for unseen testing data \cite{lu2021learning}. This foundational architecture is crucial for accurately capturing the complex relationship between microstructure and nonlinear mechanical behaviors. Second, we enhanced the DeepONet by integrating symmetries and prior assumptions directly into the network architecture as inductive biases. This integration was achieved through the equivariance-preserving unit mentioned earlier. Third, we reduced the dimensionality of the input data by extracting representative 2D cross-sections from the original 3D structures, as shown in \textbf{Fig.~\ref{fig:NN}A}). Additionally, we augmented the dataset by generating similar microstructures using the underlying sampler, while maintaining the assumption that these synthetic microstructures exhibit the same mechanical behavior as their real counterparts. These methodologies collectively empower our SciML approach to achieve robust predictive capabilities, significantly reducing the reliance on extensive training data from raw in situ experiments.

While the study here reported primarily focused on micro-scale spinodal architected metamaterials, it is important to recognize the broader applicability of the method to a wide range of other mechanical metamaterials with diverse mechanical properties. This versatility stems from DeepONet's robust capability for approximating and generalizing nonlinear function-to-function relationships. However, it is important to acknowledge that the presented workflow, particularly in the investigated scenario of limited data availability, may not be universally applicable to all base material types. The photoresist we used in this study, IP-Dip, typically exhibits smooth and reproducible stress-strain curves, as shown in \textbf{Fig. S4}. These mechanical behaviors are a key factor in the success of the approach relying on very limited experimental data. In contrast, materials with inherently more brittle characteristics, such as glassy carbon \cite{bauer2016approaching, zhang2019lightweight, wang2022achieving, guell2019ultrahigh}, or those with complex microstructures like nanocrystalline metals \cite{jin2018grain, greer2011plasticity, bernal2015intrinsic, filleter2012nucleation}, often display much more stochastic mechanical responses due to the random distribution of defects within these materials. For such materials, where mechanical responses are significantly influenced by material stochasticity, advanced SciML methods like statistical learning, Bayesian optimization \cite{yang2021b}, and uncertainty quantification \cite {psaros2023uncertainty} can be employed to extend the framework here advanced. These hybrid approaches should be capable of accounting for the inherent randomness in mechanical properties, thereby providing a more accurate model for these types of materials still with limited experimental data. Moreover, when advanced simulation tools, such as reduced-order models for stochastic metamaterials, become available, a multi-fidelity SciML model could be employed \cite{meng2020composite, lu2020extraction,howard2023multifidelity}. Such a model would leverage both cost-effective, low-fidelity simulations and expensive, high-fidelity experimental data, enabling the inverse design of a wide range of metamaterials with complex material and structural behaviors, bridging the gap between simulations and real-world experimental data.

In this study, we have developed a novel SciML model based on DeepONet to model the relationship between microstructure and the nonlinear mechanical properties of micro-architected spinodal metamaterials, utilizing a limited but high-quality experimental dataset from in situ SEM experiments on TPL-fabricated samples. We first showed that the trained DeepONet can accurately predict the nonlinear, anisotropic mechanical properties of various microstructures, including those not included in the training data. Subsequently, we leveraged the trained DeepONet for the inverse design of microstructures, aiming to achieve the desired stress-strain behaviors. Through experimental validation, we showed that the selected design is accurate for target stress-strain curves that are unseen in the training stage. These results demonstrate the effectiveness of integrating state-of-the-art SciML algorithms with advanced nano- and micro-scale fabrication and characterization techniques in designing complex metamaterials, even under data scarcity. We believe this work offers a new perspective in the field of \textit{materials-by-design}, potentially paving the way for the discovery of next-generation metamaterials with unprecedented mechanical properties.

\section*{Materials and Methods}

\subsection*{Sample fabrications}

The micro-architected spinodal structures were fabricated through a two-photon lithography system (Nanoscribe, GmbH), using the commercial two-photon photoresist (IP-DIP). The structures were printed on Silicon wafers using the dip-in mode with a 25x objective lens. To enhance the adhesion between the structures and the substrate, the Si wafers were cleaned using isopropyl alcohol (IPA) followed by de-ionized (DI) water. The writing speed and the femtosecond laser power were set as 10 $\mu$m/s and 34 mW to yield optimal printing quality. After printing, the structures were developed in propylene-glycol-monomethyl-ether-acetate (PGMEA) for 20 mins and thereafter immersed in IPA for 15 mins to remove residual PGMEA. Finally, the structures were air-dried, and no obvious structural collapse or distortion was observed. For each microstructure, we printed along three principal directions for micro-mechanical testing. The samples were uniformly coated with 8 nm Osmium using an Osmium Plasma Coater (OPC-60A, SPI) to reduce charging during SEM observation, which has negligible effects on their mechanical behaviors.

\subsection*{In situ micromechanics experiments}

Uniaxial in situ SEM compression experiments were conducted inside an FEI Nova 600 SEM, using a commercially available nanomechanical testing platform (Alemnis AG) with displacement control capability (\textbf{Fig. S5}). The displacement field was prescribed by a programmable piezo actuator paired with a 200 $\mu$m diameter flat punch. The resolution of the piezo actuator is 1 nm. The punch speed was set to maintain a quasi-static strain rate of $2 \times 10^{-3}s^{-1}$. The load was gauged using a load cell with a capacity of 1 N and 4 mN resolution. The displacement profile was corrected for the compliance of the instrumentation and substrate. Subsequently, the engineering stress-strain profiles were obtained from corrected load-displacement profiles by normalizing the samples’ footprint area and height, respectively. The energy absorption was calculated from the enclosed area between the loading and unloading curves. The ultimate stress was selected as the maximum stress during loading. The stress-strain curves are reproducible when testing on the same structure (\textbf{Fig. S6}).

\subsection*{Data Preparation}
The procedure reported in \cite{kumar2020inverse} was used to generate spinodal microstructure data. Detailed information on the data generation process can be found in \textbf{Supplementary Materials}. Briefly, the procedure allows generation of random microstructures by inputting three phase angles as control parameters. After obtaining the 3D microstructures, represented as 3D arrays $\{\Phi_{ijk}\}_{i,j,k=1,1,1}^{N,N,N}$, where $N=51$ is the number of grid points along each edge of the cubic microstructure, we reduced the dimensionality of the data by slicing the structures and extracting their cross-sections, denoted as $\{\phi_{ij}^{(pq)}\}_{p,q,i,j=1,1}^{3,n,N,N}$, where $p$ indicates the normal direction, $q$ the index of the cross-section, $i$ and $j$ the indices of the grid points, and $n=7$ the number of cross-sections in one direction. Representative cross-sections for three microstructures are plotted in \textbf{Fig. S7}. For simplicity, we omit indices $i$ and $j$ in other parts of this paper for conciseness. After generating the microstructures, we selected one sample from each set of phase angles to conduct the mechanical tests. The remaining microstructures, which serve to augment the experimental dataset, were matched to the stress-strain relationship according to their phase angles. After obtaining the paired stress-strain data from experiments, we interpolated the stress values at fixed strain intervals ($0, 0.01, 0.02, ..., 0.3$) for all samples and used them in the DeepONet training and testing.

\subsection*{DeepONet model}
The DeepONet architecture is presented in \textbf{Fig.~\ref{fig:NN}}. The branch net in our model, implemented as a convolutional neural network (CNN), processed the 2D cross-sections of the microstructures. The branch net input has 1 channel, taking an individual cross-section as its input and outputting a vector of size $m$. The $3n$ vectors (each with size $m$) were fed into the equivariance-preserving unit. They first went through an element-wise max-pooling within the three groups of $n$ vectors. This pooling is designed to achieve invariance to permutations of the cross-sections. After this step, there are three vectors, each with dimension $m$ representing the information in one spatial direction. Then, they were sorted in an element-wise way, resulting in again three  vectors as the structural encoding. The directional encoding is one of the three (unsorted) vectors, the choice of which depends on the stress component that is calculated. The four vectors were flattened into one single vector with size $4m$ and finally fed into a shallow fully connected neural network, obtaining the output vector with size $p$. The trunk net input is the combination of the loading strain $\varepsilon$ and the pseudo-time $t$ representing the loading/unloading stage. It was fed into a fully connected neural network with output size $p$. The final output of the entire DeepONet, calculated by the dot product of the two vectors with size $p$, is the stress prediction $\hat{\sigma}$ corresponding to the input microstructure $\{\phi^{(pq)}\}_{p,q=1,1}^{3,n}$, evaluated at the strain $\varepsilon$ and loading stage $t$.

To train the network, we defined the loss function as the MSE between the experimental and predicted stress-strain relationship. To accommodate the data for the neural network, we normalized the stress and strain data to be $\mathcal{O}(1)$, by $\tilde{\varepsilon}=6.67\varepsilon - 1.0$ and $\tilde{\sigma}=\sigma/\sigma_0$, where the quantities with tilde refer to the normalized values used for the DeepONet, those without the tilde refer to the original values, and $\sigma_0=50\text{ MPa}$. The learning rate was $10^{-4}$ and halved every $10^4$ epochs. The network was trained through $2\times 10^4$ epochs in total. More technical details of the DeepONet architecture are explained in the \textbf{Supplementary Materials}.

\section*{Acknowledgment}
H.D.E. acknowledges the financial support from the Air Force Office of Scientific Research (AFOSR-FA9550-20-1-0258), National Science Foundation (grant CMMI-1953806), Office of Naval Research (grant N00014-22-1-2133). G.E.K. would like to acknowledge support by the  MURI-AFOSR FA9550-20-1-0358 project, and the ONR Vannevar Bush Faculty Fellowship (N00014-22-1-2795). S.K. acknowledges the support of the Office of Naval Research through grants N00014-15-1-2935 (for acquisition of the two-photon 3D Direct Laser Writer) and N00014-23-1-2529. H.J. acknowledges the two-photon printing training from Dr. Abhishek Amrithanath.

\section*{Author Contribution}
H.J. and E.Z. initiated the collaboration and contributed equally to this work. H.D.E, G.E.K., H.J., and E.Z discussed and conceived the research idea. H.J. and B.Z. fabricated the samples and performed the in situ SEM experiments. E.Z. developed the ML algorithms. S.K. guided the two-photon printing. H.J., E.Z., H.D.E., and G.E.K. analyzed the ML outcomes. H.J. and E.Z. prepared the figures and wrote the initial draft with inputs from other authors. H.D.E, G.E.K., and S.K. supervised the work. All authors contributed to analyzing the data, revising the manuscript, and giving final approval for the publication.

\bibliographystyle{elsarticle-num-names}
\bibliography{reference.bib}
\end{document}